%% file: MR_TTV_RV.tex
\documentclass{aa}
\usepackage{amssymb,amsmath,amsfonts}
\usepackage{graphicx}
\usepackage{color}
\usepackage{ulem}             
\usepackage{multicol}
\usepackage{caption}
\usepackage{pgf,tikz}
\usetikzlibrary{arrows}
\usetikzlibrary{trees}
\usetikzlibrary{snakes}
\usetikzlibrary{shapes}
\usetikzlibrary{matrix}
\usetikzlibrary{calc}
\usetikzlibrary{positioning}
\usetikzlibrary{chains}
\usetikzlibrary{shadows}

\tikzstyle{Point} = [fill, radius=0.08]
\tikzstyle{BigPoint} = [fill, radius=0.13]
\tikzstyle{Leaf} = [color = gray]
\tikzstyle{Line1} = [dashed]
\tikzstyle{Line2} = [dotted, ultra thick]\usepackage{colortbl}

\usepackage{xcolor}

\newcommand{\be}{\begin{equation}}
\newcommand{\ee}{\end{equation}}

\usepackage[varg]{txfonts}
\usepackage{natbib}

\input{fullstrs.txt}

\input{parfullstrs.txt}
\input{controlledstrs.txt}
\newcommand\DMMRthre{$0.05$}

\begin{document}

\title{Resonant sub-Neptunes are puffier 
%
}
\titlerunning{Resonant sub-Neptunes are puffier}

\author{
Adrien Leleu$^1$,
Jean-Baptiste Delisle$^1$,
Remo Burn$^2$,
André Izidoro$^{3}$,
Stéphane Udry$^1$,
Xavier Dumusque$^1$,
Christophe Lovis$^1$,
Sarah Millholland$^{4,5}$,
Léna Parc$^1$,
François Bouchy$^1$,
Vincent Bourrier$^1$,
Yann Alibert$^6$,
Jo\~{a}o Faria$^1$,
Christoph Mordasini$^6$,
and Damien Ségransan$^1$
}
\authorrunning{A. Leleu et al}

\institute{
Observatoire de Gen\`eve, Universit\'e de Gen\`eve, Chemin Pegasi, 51, 1290 Versoix, Switzerland. \and
Max Planck Institute for Astronomy, Königstuhl 17, 69117 Heidelberg, Germany. \and
Department of Earth, Environmental and Planetary Sciences, 6100 MS 126, Rice University, Houston, TX 77005, USA. \and
Department of Physics, Massachusetts Institute of Technology, Cambridge, MA 02139, USA. \and
MIT Kavli Institute for Astrophysics and Space Research, Massachusetts Institute of Technology, Cambridge, MA 02139, USA. \and
Department of Space Research \& Planetary Sciences, University of Bern, Gesellschaftsstrasse 6, CH-3012 Bern, Switzerland.
}
\abstract
{ 
A systematic, population-level discrepancy exists between the densities of exoplanets whose masses have been measured with transit timing variations (TTVs) versus those measured with radial velocities (RVs). Since the TTV planets are predominantly nearly resonant, it is still unclear whether the discrepancy is attributed to detection biases or to astrophysical differences between the nearly resonant and non resonant planet populations. We defined a controlled, unbiased sample of \nbpcontrolled \, sub-Neptunes characterised by \textit{Kepler}, TESS, HARPS, and ESPRESSO. We found that their density depends mostly on the resonant state of the system, with a low probability (of \pcontrolledm \,) that the mass of (nearly) resonant planets is drawn from the same underlying population as the bulk of sub-Neptunes. Increasing the sample to 133 sub-Neptunes reveals finer details: the densities of resonant planets are similar and lower than non-resonant planets, and both the mean and spread in density increase for planets that are away from resonance. This trend is also present in RV-characterised planets alone. In addition, TTVs and RVs have consistent density distributions for a given distance to resonance. We also show that systems closer to resonances tend to be more co-planar than their spread-out counterparts. These observational trends are also found in synthetic populations, where planets that survived in their original resonant configuration retain a lower density; whereas less compact systems have undergone post-disc giant collisions that increased the planet's density, while expanding their orbits. Our findings reinforce the claim that resonant systems are archetypes of planetary systems at their birth.
}


\maketitle

\section{Introduction}
\begin{figure*}[!ht]
\begin{center}
\includegraphics[width=0.49\linewidth]{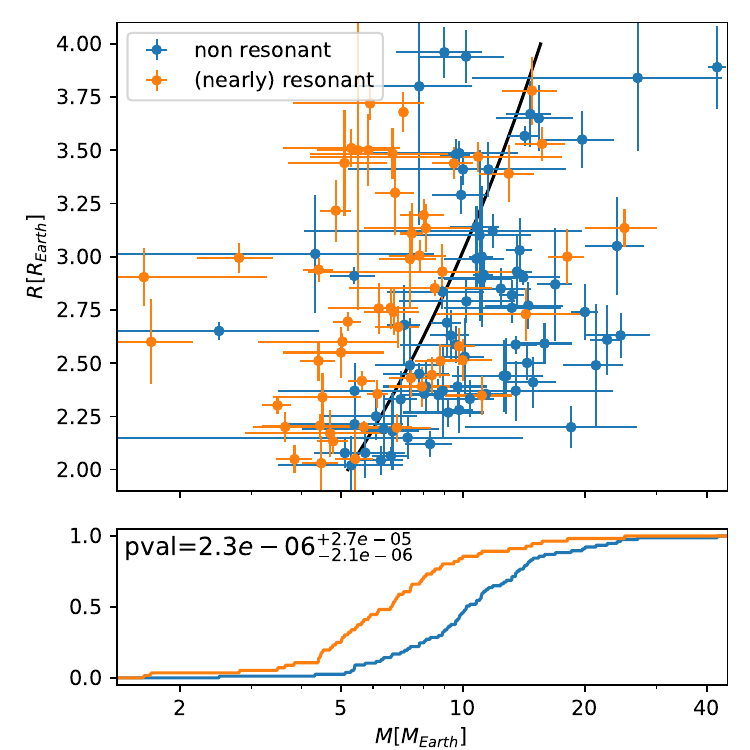}  \includegraphics[width=0.49\linewidth]{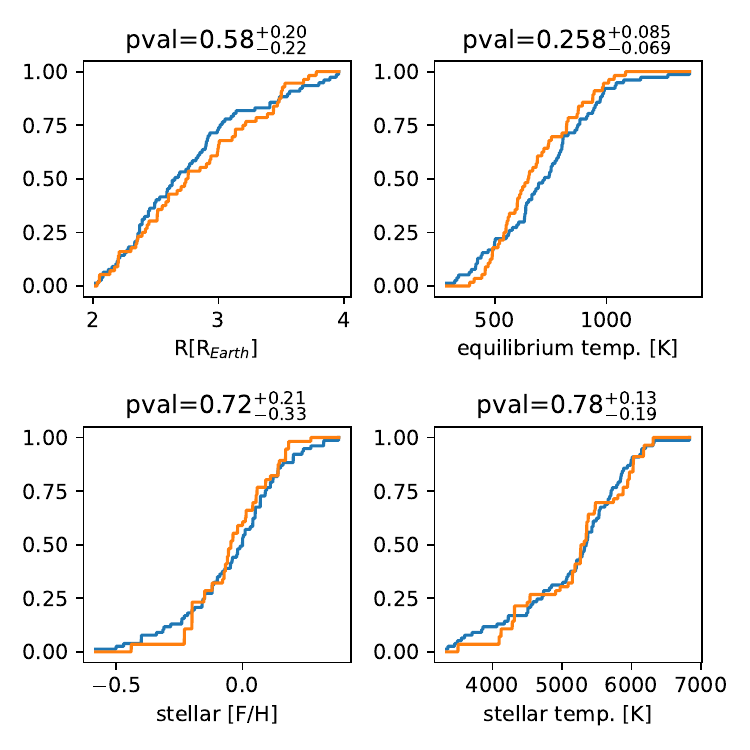}
\caption{\label{fig:full} Full sample of \nbpfull \, sub-Neptunes used in this study. The top-left panel shows the mass-radius relation. The (nearly)-resonant population is defined as  $\Delta_{MMR}<$\DMMRthre \, (see Eq. \ref{eq:DMMR}), while the non-resonant population is defined as  $\Delta_{MMR}>$\DMMRthre . The black line is the sub-Neptune mass-radius relation from \cite{Parc2024}. Other panels show the cumulative distributions for parameters of the planets or their host star. The $p_{value}$ given for each parameter is the probability that the distribution of that parameter is drawn form the same underlying distribution for the (nearly) resonant and non-resonant populations. The potential 2D correlation between parameters is explored in Fig. \ref{fig:fullcomp}.
}
\end{center}
\end{figure*}

 Planets with radius in the 1-4$R_{Earth}$ range are estimated to exist within a hundred days of orbital period around $30-50\%$ of all Sun-like stars \citep{Lovis2009,Howard2012,Fressin2013}.
In order to understand the nature of these objects, it is crucial to constrain both their masses and radii, and, thus, their densities. The bulk of exoplanet discoveries is done by transit surveys such as \textit{Kepler}/\textit{K2} and TESS, which measure the planets' radii. Then, the mass is typically estimated using the radial velocity method (RVs). For compact multi-planetary systems, the mass 
can also be estimated using transit timing variations (TTVs).
In particular, when the period ratio of two planets is close to commensurability, namely, $P_{out}/P_{in} \approx (k+q)/k$, the planets can exhibit TTVs due to their proximity to the mean motion resonance (MMR). Despite their relative rarity \citep{Fabrycky2014}, (nearly) resonant systems are over-represented in the population of planets with both mass and radius measurements because the four-year baseline of the \textit{Kepler} mission allowed estimations of their masses through TTVs at no additional cost.

Over the last decade, numerous studies \citep[e.g.][]{WuLi2013,WeissMarcy2014,Steffen2016,MillsMazeh2017,HaLi2017,Cubillos2017,Millholland2019,Leleu2023,Adibekyan2024} have noted (and discussed), the apparent discrepancy in density between the planets characterised by TTVs and radial velocities RVs. However, the origin of this discrepancy remains unclear: it could be due to sensitivity biases inherent to each method, with photometry biased towards larger planets and radial velocity biased towards more massive planets. Recent results also showed that part of the TTV-characterised population had underestimated densities due to the difficulty of extracting transit timings for low-signal-to-noise ratio (low-S/N) transits \citep{Leleu2023}. \cite{HaLi2017} put forward a selection bias as possible explanation, since TTVs tend to allow the characterisation of small planets on larger orbital periods (hence, cooler orbits) than the bulk of the RV characterisation. It has also been proposed that the systems characterised by RVs and TTVs formed in different environments, such as with different disk metallicity \citep{Adibekyan2024}. {However, the differences in physical properties could be due to the orbital configuration in which the planets are embedded \cite[e.g.][]{WeissMarcy2014,MillsMazeh2017,Goyal2023}, since TTVs mainly characterise  sub-Neptunes that are near mean motion resonances (MMRs), while the RV-characterised planets are more representative of the bulk of known exoplanets. } In this paper, we explore the possibility that there is an intrinsic connection between the densities of sub-Neptunes and their resonant orbital configurations. 

\section{Controlled sample}
\label{sec:controlled}



The population shown in Fig. \ref{fig:full} is taken from the NASA Exoplanet Archive\footnote{\url{https://exoplanetarchive.ipac.caltech.edu/}}. As of 5 March 2024, the catalogue had 695 planets for which the mass and radius, as well as the host mass, radius, effective temperature, and metallicity are given, along with their uncertainties. The host properties are required in order to check for possible correlations between these parameters and the densities of the planets. Restricting this population to close-in systems with periods in the 5-60 days range and radii between 2 and 4 $R_{Earth}$ has reduced this number to 133. The lower limit of the period range is chosen to avoid {the lower part of the Neptunian desert, which is partly shaped by photoevaporation \citep{Owen2018}.}
We define the (nearly) resonant population (in orange) as planets whose period ratio with an inner or outer planet satisfies $\Delta_{MMR}<$\DMMRthre , where 

\be
\Delta_{MMR} = \left| \frac{P_{out}}{P_{in}}  - \frac{k+q}{k}\right|
\label{eq:DMMR}
,\ee
for $q=1$ and $k\in [1,2,3,4,5]$ or $q=2$ and $k\in [3,5]$. While the non-resonant population (in blue) is defined $\Delta_{MMR}>$\DMMRthre . This limit is set by the edge of the clump of nearly-resonant system found in \textit{Kepler} \citep{Fabrycky2014}. 
The (nearly) resonant population, in orange, appears to be composed of lower-density planets than the non-resonant population. However, these populations could be affected by numerous biases. Notably, the {(nearly-)resonant} population is mainly characterised by TTVs, while the {non-resonant} population is mainly characterised by RVs.

\begin{figure*}[!ht]
\begin{center}
\includegraphics[width=0.49\linewidth]{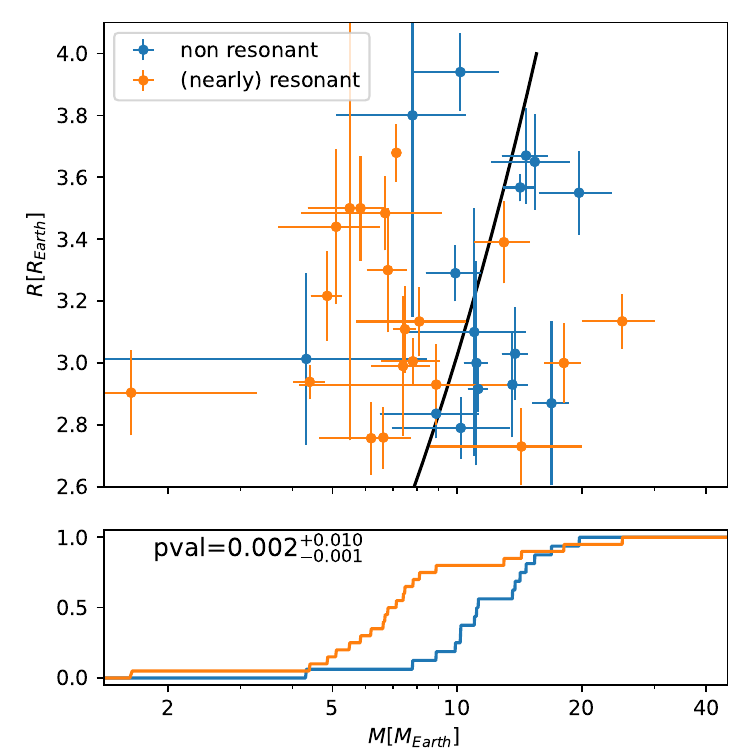}
\includegraphics[width=0.49\linewidth]{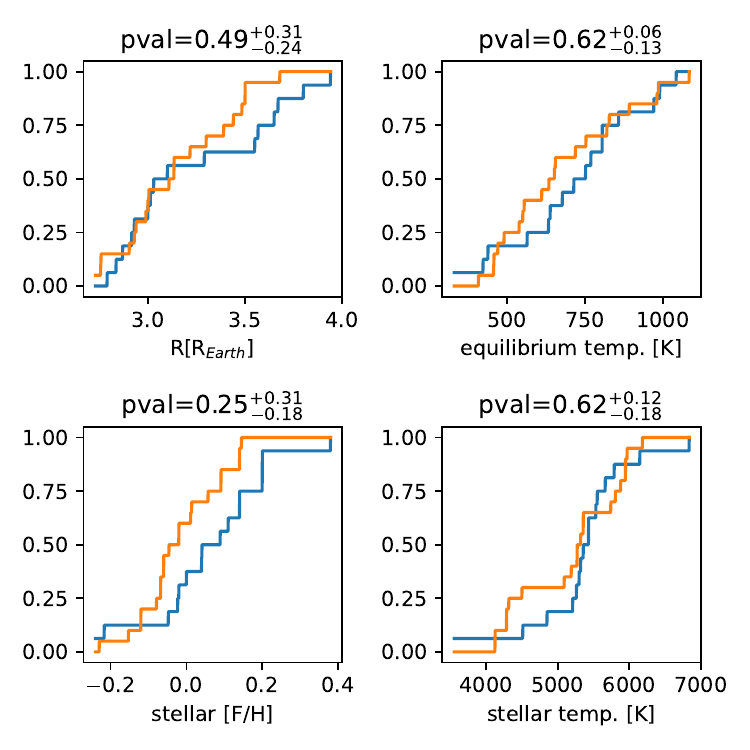}  
\caption{\label{fig:controlled} Same as Fig. \ref{fig:full}, for the controlled sample.
 The potential 2D correlation between parameters are explored in Fig. \ref{fig:controlled_fullcomp}.
}
\end{center}
\end{figure*}

{
Regarding the RV-characterised masses, a possible bias could come through the follow-up observation process; 
some planets could be dropped out 
after few RV points were taken,
if the RV signature of the planets did not seem large enough.
This selection process would lead to a bias in the literature towards higher masses and higher densities for RV-characterised planets.
Using all sub-Neptunes in the 5-60 days range that were followed-up by HARPS or ESPRESSO, we show in Appendix \ref{ap:controlled} that such bias is absent for planets whose radius is above 2.7R$_{Earth}$, since 93\% (26 out of 28) of the followed-up planets in that radius range have published masses. These 20 RV-characterised planets are the first part of our controlled sample.

For TTV-characterised planets, the mass-radius relation can strongly be affected by mass-eccentricity degeneracies \cite{Lithwick2012} and the manner in which the TTVs are extracted from the light curves \citep{Leleu2023}. To address the first point, we only use planets whose mass estimations have been shown to be robust against mass and eccentricity degeneracy \citep[e.g.][]{HaLi2017,Leleu2023}.
Regarding the second point, \cite{Leleu2023} showed that TTVs are typically correctly estimated by usual methods if the signal to noise ratio of individual transits (S/N$_i$) is high enough. 
As shown in Fig. \ref{fig:SNRi}, we checked that the planets with radius above 2.7 had an S/N$_i > 3.5$, which ensures that their individual transit timing can robustly be recovered \citep{Leleu2023}. 
Here also, it translates to the selection of stars that are bright enough and not overly active. If individual transits of planets can reliably be observed, there is no reason why we could not characterise denser planets, as their TTV signals would either be larger \citep{Lithwick2012} or faster \citep{NeVo2016}, depending on whether the pair is near or inside a MMR. 

We therefore define our controlled sample as the planets in the 2.7 to 4 $R_{Earth}$ range in the mass-radius diagram (shown in Fig. \ref{fig:controlled}). This sample, detailed in Table \ref{tab:controlled} is made of \nbpcontrolledres{} (nearly) resonant planets and \nbpcontrollednres{} non-resonant planets. Using the Kolmogorov-Smirnov test, we estimated that the radius distributions of these two populations have a $p_{value}=$\pcontrolledr \, probability\footnote{Median and uncertainties on the $p_{value}$ are estimated by drawing 1000 samples assuming a Gaussian distribution for the radius of each planet, then computing the 0.16, 0.5, and 0.84th quantiles of the resulting $p_{value}$ distribution.} to be drawn from the same underlying population. However, the probability that their masses have been drawn from the same underlying population is $p_{value}=$\pcontrolledm. We then checked whether this discrepancy could be attributed to different stellar metallicities, equilibrium temperatures \citep[e.g.][]{HaLi2017}, or stellar effective temperatures. With respect to all of these quantities, the two populations are similar, with $p_{value}$ of \pcontrolledteq, \pcontrolledmet \,, and \pcontrolledteff \,, respectively. Exploring possible 2D relations between these parameters, we performed 2D Kolmogorov-Smirnov tests \citep{Peacock1983}. These results are shown in Appendix \ref{ap:controlled}. Across all our tests, the $p_{value}$ involving the mass are lower than the rest by two orders of magnitudes. We therefore conclude that the proximity to MMR is the main factor in the discrepancy between the mass, hence, the density, of the two sub-populations.

\section{Full sample}
\label{sec:full}


We go on to consider the full sample shown in Fig. \ref{fig:full}, taking the robust masses from \cite{HaLi2017} and \cite{Leleu2023} when available. For this sample, the probability that the mass of the (nearly) resonant and non-resonant population is drawn form the same underlying population drops to $p_{value}=$\pfullm \,, while the rest of the explored parameters are consistent between the two sub-population. The full 2D comparison is in Appendix \ref{fig:fullcomp}. 
The 2D comparison was also performed for the full sample restricted to the 2.7-4$R_{Earth}$ (see Fig. \ref{fig:parfullcomp}).


\begin{figure*}[!ht]
\begin{center}
\includegraphics[width=0.61\linewidth]{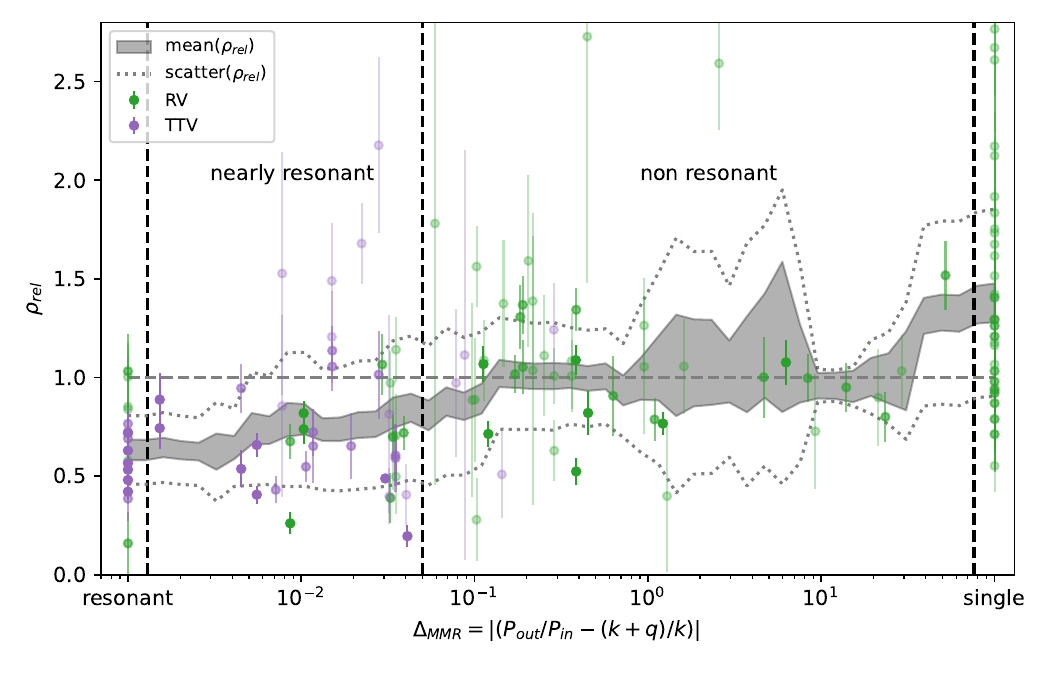}\includegraphics[width=0.388\linewidth]{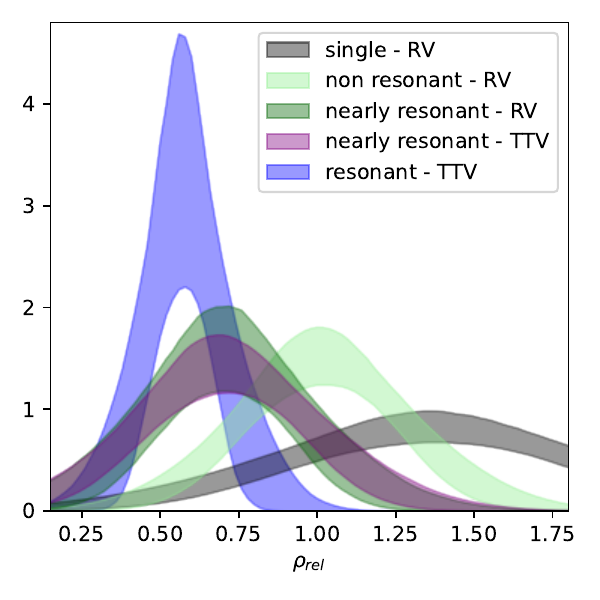}
\caption{\label{fig:DMMRrho} Relative density of planets as a function of the distance to the closest MMR (left). Resonant chains and single planets are arbitrarily set at $\Delta_{MMR}=0.001$ and $100$, respectively. 
The colors indicate the method used to obtain the mass. The dark grey area indicates the 1$\sigma$ confidence interval for the local mean value of $\rho_{rel}$, while the dotted lines show a local estimation of its scatter. 
\textit{} Distribution of relative densities assuming gaussian distributions, binning by method and distance to the resonance (right): resonant, nearly resonant ($0.001<\Delta_{MMR}<0.05$), non resonant ($0.05<\Delta_{MMR}$), and single planets.    
}
\end{center}
\end{figure*}

To compare the relative densities of planets with different radii, we defined $\rho_{rel} = \rho / \rho_{ref}$, which is the ratio between the density measured for a planet and a reference density for a planet of the same radius, using the M-R relation from \cite{Parc2024} (black line in Fig. \ref{fig:full}). To further illustrate the effect of the proximity to MMR on the density of planets, in Fig. \ref{fig:DMMRrho} we show $\rho_{rel}$ as a function of $\Delta_{MMR}$. Formally resonant planets (including resonant chains) are shown on the left, while single planets are shown on the right\footnote{The position of single planets in this figure is arbitrary, and there are possibly non-transiting planets in their system. However, given the relative rarity of planets near MMRs, we assume that they are part of the non-resonant population.}. 
In the figure, we estimated the local mean and scatter of $\rho_{rel}$ by fitting a Gaussian distribution in a box sliding over $\log_{10}(\Delta_{MMR})$ with a width of $1$.
Most planets for which $\Delta_{MMR}<0.05$ have a relative density below 1, and this is even more so for the formally resonant systems, which have a mean relative density estimated at $0.63 \pm 0.05$. On the contrary, planets for which $\Delta_{MMR}>0.1$ are more uniformly spread around $\rho_{rel}=1$ and single planets are on average denser, with a mean relative density of $1.38 \pm 0.10$.
The right panel shows the envelope of the $0.16-0.84$ quantiles of the Gaussian distributions of the relative density as a function of the mass measurement method used and the distance to the resonance. Only the nearly resonant population ($0.001<\Delta_{MMR}<0.05$) had enough measurements to be estimated by both methods. From this analysis, we have drawn three observations. First, the correlation between the distance to MMR and the relative density of sub-Neptunes is visible when using RV-characterised planets alone (grey, green, and dark green Gaussians). Second, RV- and TTV- characterised planets have similar relative density where they overlap in the nearly-resonant population (dark green and dark purple Gaussians). Third, the distribution of relative densities increase both in mean value and in spread away from the resonance, with resonant planets having similar relative densities, while single planets have a larger spread. 


\section{Discussion}
\label{sec:discussion}
\begin{figure}[!ht]
\begin{center}
\includegraphics[width=0.99\linewidth]{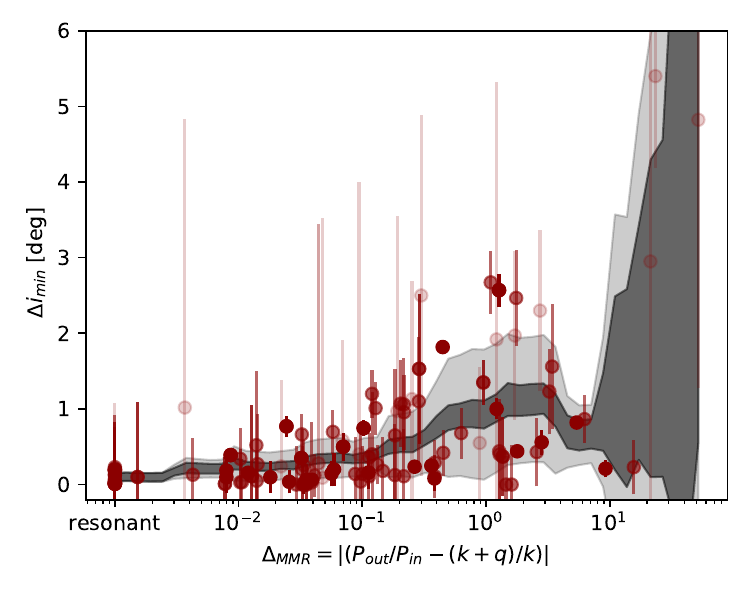}
\caption{\label{fig:DMMRDi} Minimal mutual inclination of transiting confirmed planets. The dark grey area indicates the 1$\sigma$ confidence interval for the local mean value of $\Delta i_{min}$, while the lighter grey area shows a local estimation of its scatter.} 
\end{center}
\end{figure}

\begin{figure*}[!ht]
\begin{center}
\includegraphics[width=0.99\linewidth]{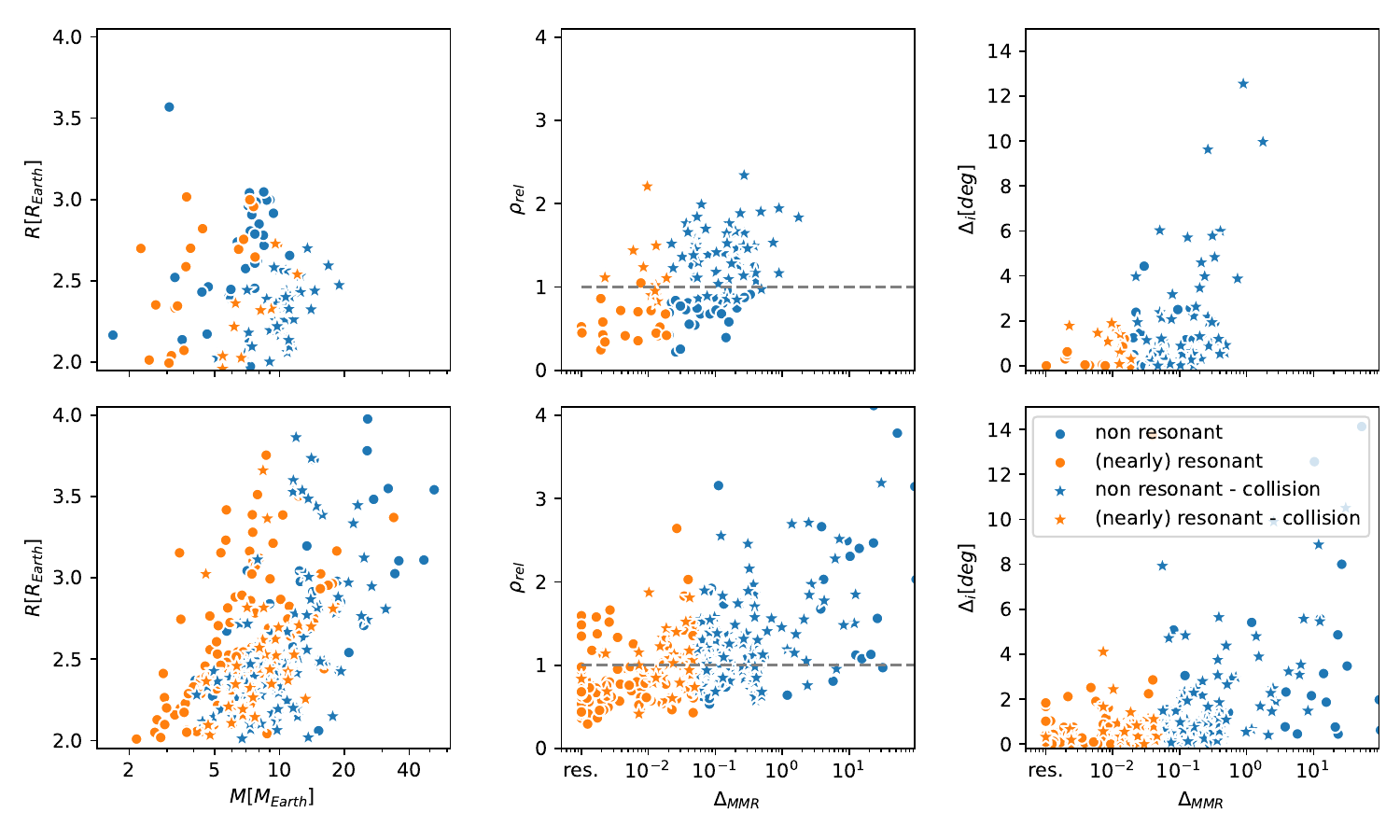}
\caption{\label{fig:models} Synthetic populations of close-in systems. Top: Population from \cite{izidoroetal21,Izidoro2022} bottom: Population form NGPPS \cite{Burn2024}.  } 
\end{center}
\end{figure*}



If the difference in density between (nearly) resonant and non-resonant planets is not due to observational or selection biases, nor to the type of star that the planets orbit, it must be a result of different formation and evolution pathways. For example, the (nearly) resonant planets could be puffier as a result of atmospheric inflation due to tidal heating \citep{Millholland2019, Millholland2020}. This would arise if the (nearly) resonant have systematically larger eccentricities or planetary obliquities. This is supported by previous studies, which suggested that planets captured in mean-motion resonances might have their spin-axes tilted as a result of secular spin-orbit resonance capture during the orbital migration process \citep{Millholland2019,Millholland2024}.   

Another hypothesis is that resonant and non-resonant planets were formed in different locations with diverse formation conditions. For example, \cite{Lee2016} posited that most super-Earths and sub-Neptunes formed in situ in gas-poor discs towards the end of the disc lifetime. However, they suggested that the rarer class of low-density ``super-puff'' planets formed outside $\sim 1$ AU and accreted thicker gaseous atmospheres due to more efficient cooling. Because they formed further out, they would have entered MMRs upon inward migration.  

Alternatively, lower densities for planets in MMRs is expected by the model known as `breaking the chains' \citep[see section 4 of][]{Bean2021}.
In that model, close-in systems of sub-Neptunes form in resonant chains due to the migration of planets in the protoplanetary discs and the positive torque at the inner edge of the disc \citep[e.g.][]{GoTre1979,Weidenschilling1985,Masset2006,TePa2007}. 
Resonant chains can then become dynamically unstable after the gaseous disc dissipates \citep{TePa2007,OgIda2009,Cossou2014}.
This model reproduces the observed period ratio and multiplicity distribution of the close-in sub-Neptune population if $\approx 95\%$ of the chains become unstable after the disc dispersal \citep{Izidoro2017}. Some of these instabilities lead to giant impacts which can eject part, or all, the primordial H/He atmospheres of the planets \citep{BiSch2019}, resulting in non-resonant planets that are on average denser than their resonant counterpart. 

This scenario leads to a second observable: known resonant chains tend to be remarkably coplanar \citep[e.g.][]{Agol2020,Leleu2021}, while planet-planet scattering is expected to increase the mutual inclination between planets. In Fig. \ref{fig:DMMRDi} we show the minimum mutual inclination (i.e. assuming that the ascending nodes of all the planets are aligned in the sky) of all pairs of planets that were confirmed in the exoplanet archive. Here, we can also see that pairs further away from MMRs have a larger scatter in the mutual inclination, in agreement with the 'breaking the chains' model. We note that the actual trend might be stronger, since here we only measure the minimal mutual inclination, the impact parameter of (nearly) resonant systems could be biased by unaccounted TTVs \citep{Garcia2011} and misaligned, spread-out systems have a lower transit probability.  

The 'breaking the chains' mechanism \citep[e.g.][]{Izidoro2017,Izidoro2022} is also observed in different planet formation models \citep[e.g. NGPPS][]{Emsenhuber2021,Burn2024}.
In Fig. \ref{fig:models}, we show synthetic systems from \cite{izidoroetal21,Izidoro2022} and NGPPS \citep{Emsenhuber2021,Burn2024}. These populations (described in Appendix \ref{ap:Andre} and \ref{ap:Remo}, respectively) simulate the formation of planetary systems from planet embryos in the proto-planetary discs up to $\sim$50 millions of years after the disc dispersal. As can be seen in the middle panels, both populations harbour lower-density planets for the (nearly) resonant sub-population, while the non-resonant populations have larger and more diverse relative densities. The right panels show that the mutual inclination between each planet pairs tend to be larger for larger distance to MMRs. In both populations, these features are linked with post-disc instabilities and collision, shown with star markers in Fig. \ref{fig:models}. These instabilities can be due to the configuration of the chain itself, but also to the existence of an outer more massive planet \citep[see Appendix \ref{ap:Remo} and also][]{Schlecker2021,Izidoro2022}. In addition, for the NGPPS population, the larger density of the non-resonant population is partially due to the accretion of more rocky embryos from the inner system during the giant impact stage after gas disk dissipation.


\section{Conclusion}
\label{sec:conclusion}

Our results support the idea that the apparent discrepancy between TTV- and RV-characterised planets is astrophysical and due to different formation and evolution pathways of the characterised populations, rather than a method-related bias. The significance of the controlled sample ($p_{value}$ of \pcontrolledm) can be improved by a homogeneous analysis of the RV-characterised population, and a larger completion in the publication of mass measurement for smaller planets. For the TTVs, the population needs to be systematically checked for the mass or eccentricity degeneracies \citep{HaLi2017} and the robustness of the TTV extraction or the photo-dynamical analysis \citep{Leleu2023}, as well as ensuring that small dense resonant planets are not missed due to large TTVs \citep{RIVERS1,RIVERS2}. In this study, we were also able to show that TTV and RV characterised planets had a similar relative density distribution for the nearly resonant population. A next step would be to get enough RV characterised resonant systems of sub-Neptunes, such as resonant chains, to check whether these results hold for that population. Finally, PLATO will enable the discovery and characterisation of systems by both TTVs and RVs, which should help to further alleviate the biases inherent to each method.

\bibliographystyle{aa}
\bibliography{biblio}

\begin{acknowledgements}
The authors acknowledge support from the Swiss NCCR PlanetS and the Swiss National Science Foundation. This work has been carried out within the framework of the NCCR PlanetS supported by the Swiss National Science Foundation under grants 51NF40$_{}$182901 and 51NF40$_{}$205606. 
AL acknowledges support of the Swiss National Science Foundation under grant number  TMSGI2\_211697.
R.B. acknowledges the support from the German Research Foundation (DFG) under Germany’s Excellence Strategy EXC 2181/1-390900948, Exploratory project EP 8.4 (the Heidelberg STRUCTURES Excellence Cluster). This project has received funding from the European Research Council (ERC) under the European Union's Horizon 2020 research and innovation programme (project {\sc Spice Dune}, grant agreement No 947634). 
\end{acknowledgements}

\begin{appendix}

\section{Controlled sample}
\label{ap:controlled}

To test for the RV selection bias, we used the ESO public archive\footnote{\url{https://archive.eso.org}} to get the number of HARPS and/or ESPRESSO (outside of GTO) measurements
of all planets with radius between 2 and 4 $R_{Earth}$ and period between $5$ and $60$ days.
We restricted our analysis to HARPS and ESPRESSO because the ESO archive allows for a query of the number of measurements for all observed targets.

\input{controlled_table.txt}
We then checked using the exoplanet archive whether these planets had a  published mass in the literature.
We considered targets that had at least 20 RV points taken by HARPS and/or ESPRESSO by March 2022.
We chose 20 points as a threshold because some targets might have been dropped early because of stellar activity or another reason not related to the mass of the planet.
We only looked at measurements taken before March 2022 to only consider planets for which the observers had enough time to analyse the data and publish the mass.
As we can see in Fig. \ref{fig:nb_equiv}, planets whose radius is below 2.7 $R_{Earth}$ might be affected by a selection bias, as 24\% of these have not been published. On the other hand, only 7\% of the planets with a radius above the 2.7 $R_{Earth}$ threshold have not been not published. 
There are also three planets with radius above  2.7 $R_{Earth}$ for which a mass is given on the archive, but no errorbar: CoRoT-24 b ($Vmag=15.38$), HIP 94235 b ($Vmag=8.3$), and K2-290 b ($Vmag=11.11$, triple star system). For CoRoT-24 b, \cite{Alonso2014} reported a 1$\sigma$ upper limit at 5.8$M_{Earth}$, but could not exclude that the planet was orbiting another star. For HIP 94235 b and K2-290 b, \cite{Zhou2022} and \cite{Hjorth2019} reported 3$\sigma$ upper limit at 379$M_{Earth}$ and 21$M_{Earth}$, respectively; however, the authors did note that their dataset did not have the precision to reach the expected mass range for these planets. Therefore, we did not include these planets in our analysis. 

Finally, we checked whether this sub-sample could have been pre-selected before their observation by HARPS or ESPRESSO. To do so, we checked the reference of each planet to see if points were taken by other facilities before their monitoring by one of these telescopes. We found only 4such instances out of the 20 planets (see notes in Table \ref{tab:controlled}). We therefore consider this sub-sample to be representative of the underlying population of planets between 5 and 60 days of the orbital periods and radii in the [2.7-4]  $R_{Earth}$ range. The controlled sample is described in Table \ref{tab:controlled}. Possible 2D correlations between the planetary and/or stellar parameters of the two populations are explored in Fig. \ref{fig:controlled_fullcomp}.

\begin{figure}[!ht]
\begin{center}
\includegraphics[width=0.99\linewidth]{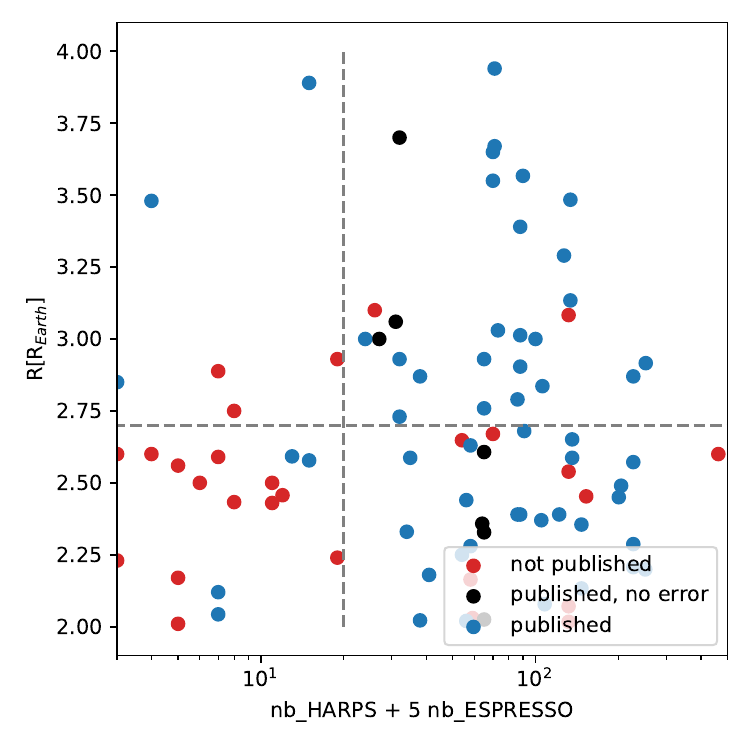}
\caption{\label{fig:nb_equiv} Number of points taken by Mars 2022 by HARPS and ESPRESSO facilities for planets in the 2 to 4 $R_{Earth}$ range. Each ESPRESSO measurements are counted as five HARPS measurements to account for photon noise. The horizontal grey dashed line represents the radius cut-off above which we consider that the published mass-radius population is not affected by selection biases.
}
\end{center}
\end{figure}

\begin{figure}[!ht]
\begin{center}
\includegraphics[width=0.99\linewidth]{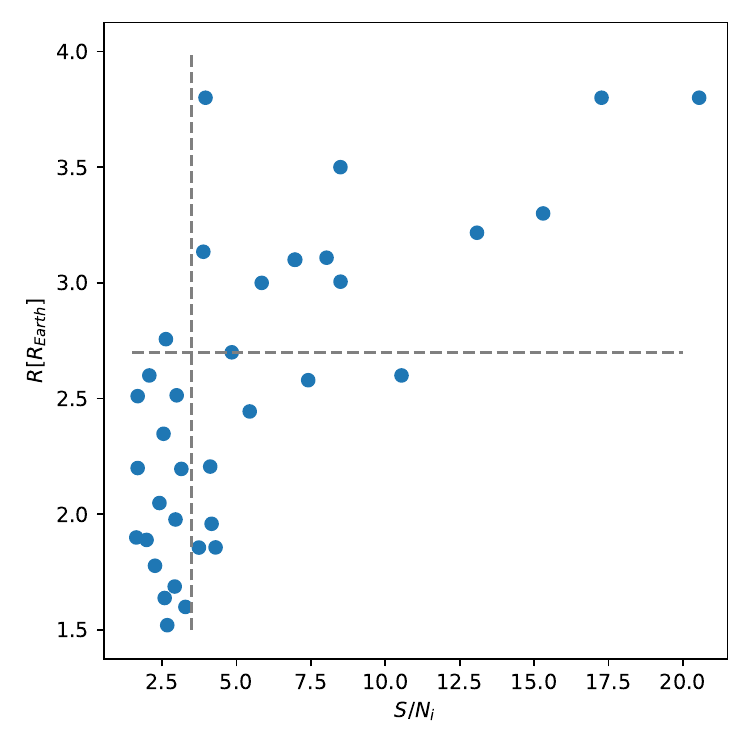}
\caption{\label{fig:SNRi} 
Radius of the \textit{Kepler} TTV characterised planets the full sample that came either from \cite{HaLi2017} or \cite{Leleu2023}, as function of the S/N of individual transits ($S/N_i$). The vertical grey dashed line shows the $S/N_i=3.5$ threshold above which we consider that large TTVs do not prevent the detection of the planet \citep{Leleu2023}. The horizontal grey dashed line represent the radius cut-off above which we consider that the published mass-radius population is not affected by selection biases. }
\end{center}
\end{figure}

\begin{figure*}[!ht]
\begin{center}
\includegraphics[width=0.99\linewidth]{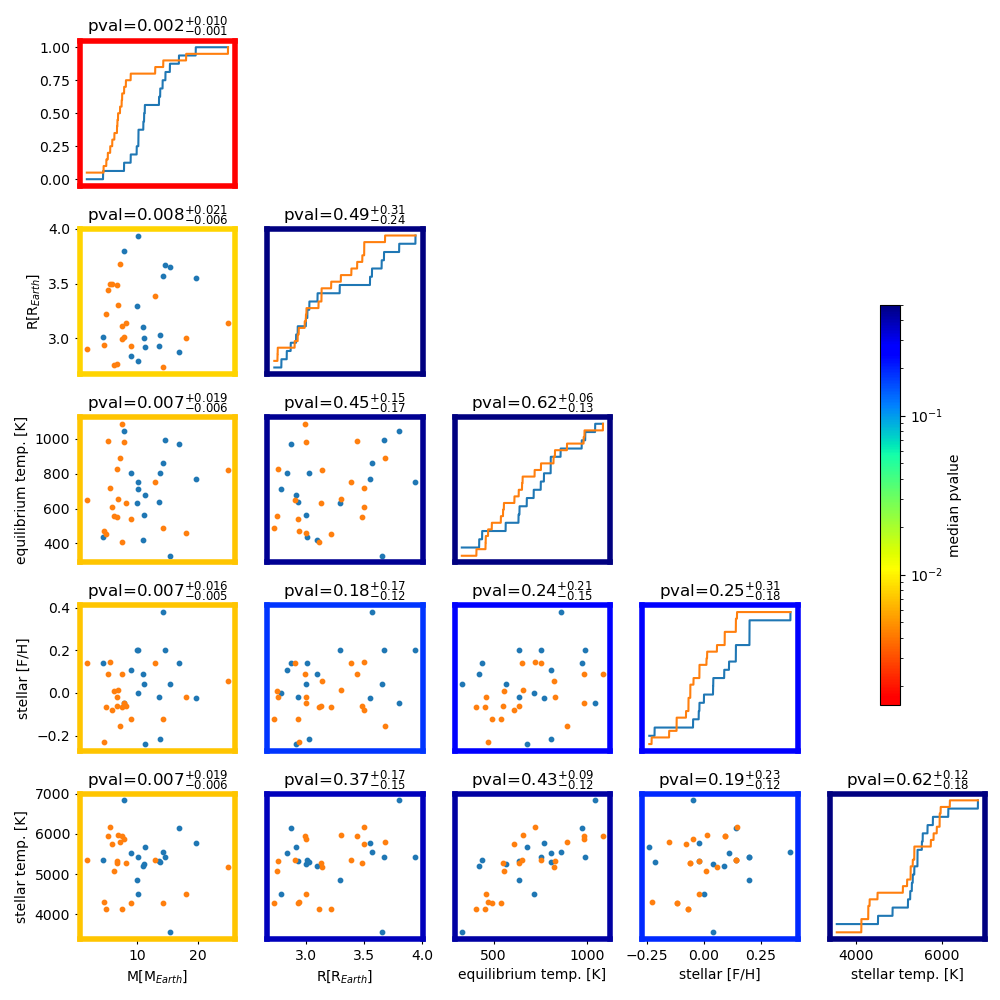}
\caption{\label{fig:controlled_fullcomp} 
Controlled sample for planets in the 2.7 to 4 $R_{Earth}$ range. (Nearly-)resonant planets are shown by orange points, while non-resonant planets are shown by blue points. The $p_{value}$ shown on the diagonal are 1D Kolmogorov-Smirnov test, while the $p_{value}$ on the bottom-left triangle are 2D Kolmogorov-Smirnov tests \citep{Peacock1983}. Median and uncertainties on the $p_{value}$ are estimated by drawing 1000 samples assuming a Gaussian distribution for the radius of each planet, then computing the 0.16, 0.5, and 0.84th quantiles of the resulting $p_{value}$ distribution.}
\end{center}
\end{figure*}

\section{Full sample}
Possible 2D correlations between the planetary and/or stellar parameters of the two populations are explored in Fig. \ref{fig:fullcomp} for the full sample given in Fig. \ref{fig:full}. The same analysis is also performed by restricting that sample to the 2.7 to 4 Earth radii range (see Fig. \ref{fig:parfullcomp}).

\begin{figure*}[!ht]
\begin{center}
\includegraphics[width=0.99\linewidth]{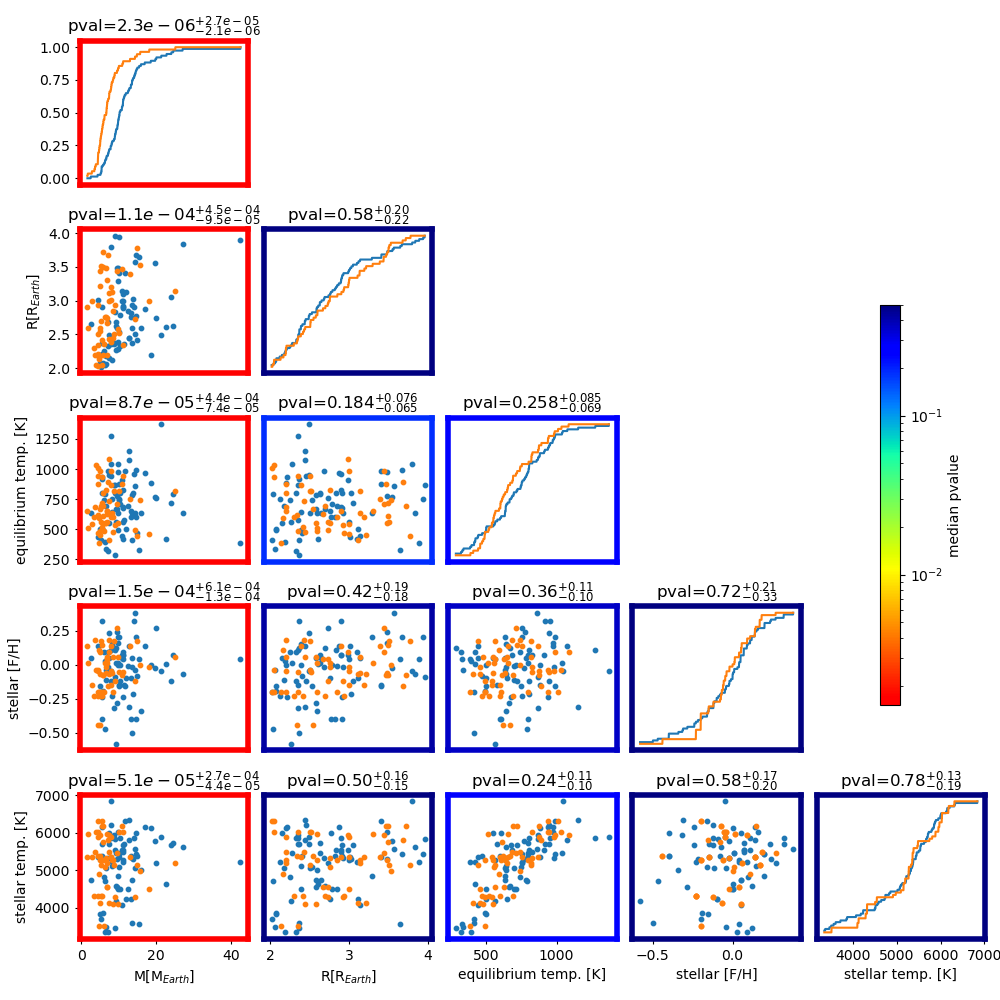}
\caption{\label{fig:fullcomp} 
Same as Fig \ref{fig:controlled_fullcomp} but for the full sample shown in Fig. \ref{fig:full}, for radius between 2 and 4 Earth radii.
}
\end{center}
\end{figure*}

\begin{figure*}[!ht]
\begin{center}
\includegraphics[width=0.99\linewidth]{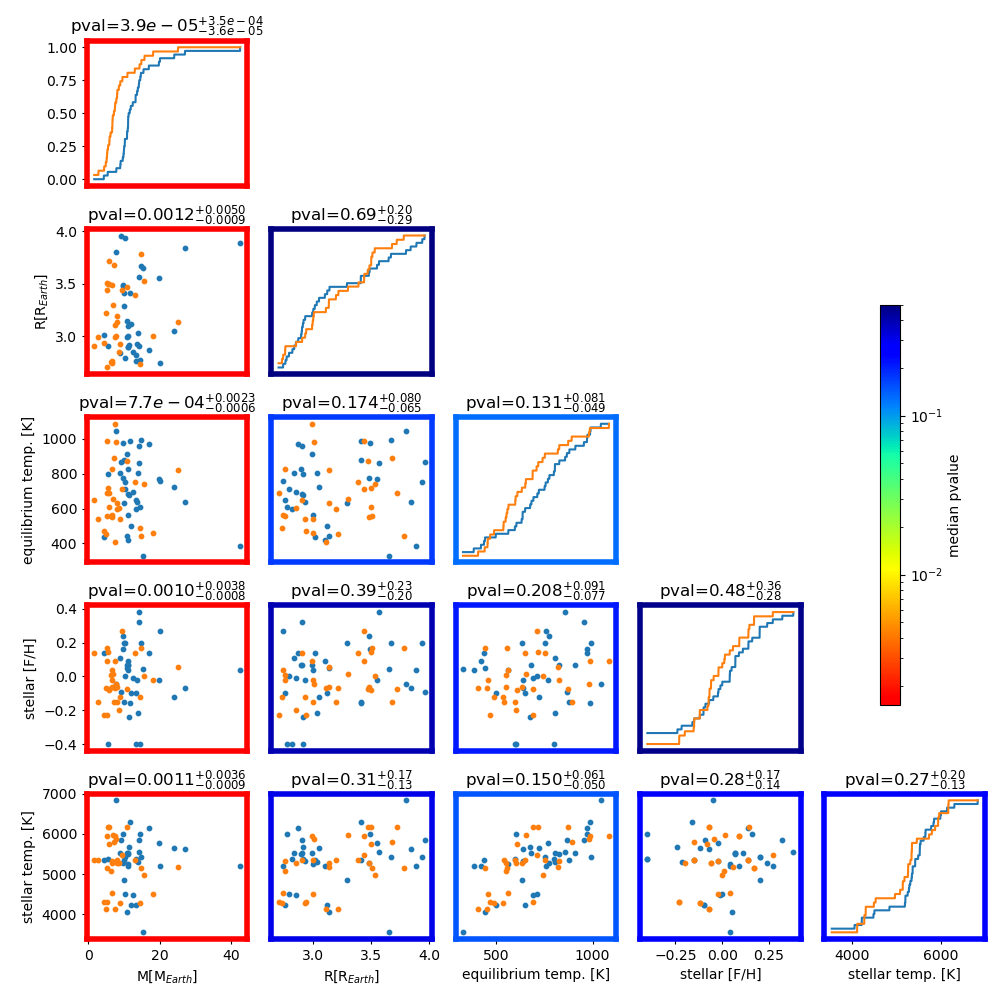}
\caption{\label{fig:parfullcomp} 
Same as Fig \ref{fig:controlled_fullcomp}, but for the full sample shown in Fig. \ref{fig:full}, for a radius between 2.7 and 4 Earth radii.
}
\end{center}
\end{figure*}


\section{Synthetic planetary populations}
\subsection{Breaking the chains model \citep{Izidoro2017}}
\label{ap:Andre}

Our initial set of simulated planetary systems originates from \cite{Izidoro2022}, building upon prior simulations of the 'breaking the chains' scenario \citep{Izidoro2017,izidoroetal21}. These simulations track the formation of super-Earths and sub-Neptunes by tracing the evolution of Moon-mass planetary seeds within a circumstellar disk. Various physical processes are considered, including gas-assisted pebble accretion \citep[e.g.][]{lambrechtsjohansen12}, gas-driven planet migration \citep[e.g.][]{baruteauetal14}, gas tidal damping of orbital parameters \citep[e.g.][]{cresswellnelson08}, and mutual gravitational interactions among planetary embryos.

In \cite{izidoroetal21}, planetary seeds grow through pebble accretion and mutual collisions. Pebbles beyond the snowline are presumed to contain 50\% water ice mass. As these pebbles migrate inward and cross the water snowline, they sublimate, losing their water component and releasing silicate grains.  Collisions are modelled as perfect merging events conserving mass and momentum.

The planet formation simulations of \cite{izidoroetal21} provide planetary mass and composition but not planet size or radius. To compare the simulations outcome with observational trends involving planet sizes such as the exoplanet radius valley and the peas-in-a-pod trend, \cite{Izidoro2022} employ mass-radius relationships \citep{Zeng19} to convert mass into planetary radius. In addition,  in this model, giant impacts (with projectile-to-target mass ratios greater than 0.1) occurring after gas disk dispersal strip primordial atmospheres, leaving behind either bare rocky or water-rich cores \citep{bierstekeretal19}. According to \cite{Izidoro2022}, approximately 80-90\% of late impacts qualify as giant impacts in their model. Their model does not account for the formation of secondary/outgassed atmospheres but for only for primordial atmospheres accreted during the gaseous disk phase (H/He rich).

The stability of a primordial atmosphere for a planet that did not experience giant impacts after gas dispersal is estimated using an energy-limited escape prescription, considering stellar X-ray and ultraviolet radiation \citep{OwWu2017}. The criterion by \cite{misenerschlichting21} compares atmospheric binding energy to the energy received by the planet from 100 million to 1 billion years. An energy ratio smaller than unit indicates sufficient energy for atmosphere photo-evaporation. Planet sizes are computed following different planet models, which account, or not, for the presence of a primordial atmosphere following the atmospheric instability criterion of \cite{misenerschlichting21}.

\subsection{New generation planetary population synthesis (NGPPS)}
\label{ap:Remo}
The second set of theoretical planetary system calculations was obtained from the planetary population synthesis exercise conducted by \citet{Emsenhuber2021b} and updated by \citet{Burn2024}. This set of simulations use the Bern model of global planet formation and evolution \citep{Emsenhuber2021}. The starting point of the simulations are protoplanetary dust and solid disks around Solar-type stars with observation-informed distributions of initial conditions (\citealp{Tychoniec2018}, see also the discussion in \citealp{Emsenhuber2023}). The gas disk viscous surface density evolution equation \citep{Pringle1981} is solved assuming an $\alpha$-viscosity of $2\times 10^{-3}$, photoevaporative mass loss, and a consistent temperature structure using opacities from \citet{Bell94}, viscous and irradiation heating. Apart from 1\% dust used as opacity source, the solid disk is assumed to be present in the form of planetesimals that are modelled as fluid and accreted by 100 growing seed embryos randomly distributed in the disk at initialisation. The composition of gas, embryos and planetesimals, initialised following \citet{Marboeuf}, is tracked during planet growth. Gas accretion onto the planets proceeds by cooling and contraction of the previously accreted gas which is modelled in one dimension assuming hydrostatic equilibrium and energy release at the boundary between the solid core and the gaseous envelope. The simulations explicitly model gravitational interactions between the embryos using the \texttt{mercury} code \citep{Chambers1999} where an additional force is added due to the interaction of the planets with the gaseous disk. The force is derived from prescriptions of type I \citep{pdk10,pdk11} and II \citep{Dittkrist14} migration timescales, as well as eccentricity and inclination damping \citep{ColemanNelson14}. This will commonly lead to an inwards movement at moderate excitation of the planets during the gas disk stage. If two embryos collide, we assume a perfect merging of the solid core (including volatile species) and immediate ejection of the hydrogen-helium envelope of the smaller of the two planets. The impact energy of the impacting core is then added as a luminosity source over a smoothing timescale on the core-envelope boundary of the larger target, which can lead to radius inflation and gas loss, namely, impact stripping. For a full description of the technical implementation, we refer to \citet{Emsenhuber2021}. Here, we note that the perfect merging of volatiles, such as water, is a current model shortcoming and impact stripping of volatiles need to be included in future versions of the model. Nevertheless, the stripping of heavier water or carbon-bearing species is less efficient than that of the lighter hydrogen and helium \citep{Burger2020}.

The evolution of the 1000 NGPPS planetary systems, following 100\,Myr of N-body integration with the aforementioned model, was re-calculated by \citet{Burn2024} considering an improved equation of state for water \citep{Haldemann2020} for each simulated planet individually. If water or other ices were present from the formation modelling, they are mixed with any H/He left at this stage. For simplicity, all volatile species are modelled as water. The revised internal structure modelling results in significantly different, larger, radii of volatile-rich planet due to the lower density of supercritical water. Over typically 0.1\,Gyr timescales, the loss of the mixture is calculated using a mass-weighted mass loss rate of the two constituents due to high-energy irradiation by the star \citep{kubyshkina2018,kubyshkina2021,Johnstone2020}. The fraction of elements in the envelope different from hydrogen and helium, $Z_{\rm env}$, is kept constant motivated by numerical results in the regime of efficient mass loss \citep{Johnstone2020}. Here, we show the planetary population at an evolutionary age of 5\,Gyr after star formation; however, we caution that due to computational limitations, the dynamical state represents a 100\,Myr-old system and, in some cases, slightly varying masses (prior to atmospheric mass loss).

\begin{figure*}
    \centering
    \includegraphics[width=\linewidth]{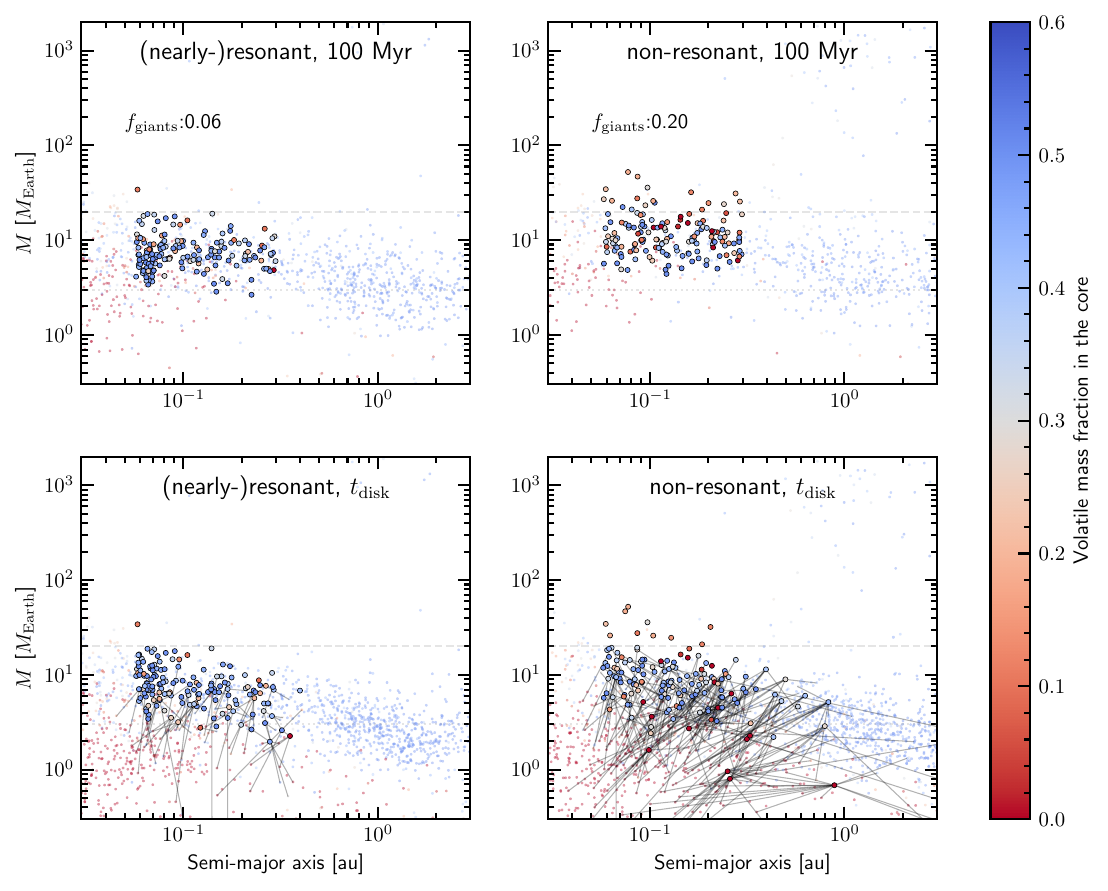}
    \caption{Simulated planetary systems from NGPPS hosting final planets comparable to the observed sample. The criteria are a radius between 2 and 4\,R$_{\oplus}$ and an orbital period between 5 and 60 days. Those planets are marked in the semi-major axis versus total mass plots (left four panels) with larger, dark outlined circles. The rest of the simulated planetary systems containing those planets is shown with transparent circles using the same color-code showing the mass of volatile species accreted as ices compared to the total core mass. The fraction of systems with giants $f_{\rm giants}$ differs between the two samples. The lower two panels on the left show an earlier stage of the system at the time of disk dispersal. In addition, collision partners of the planets meeting the selection criteria are connected to them with lines. For visual guidance, the mass region from 3 to 20\,$M_{\rm Earth}$, where most of the (nearly) resonant planets reside, is marked with grey lines.}
    \label{fig:NGPPS_a_m_scatter}
\end{figure*}

\begin{figure}
    \centering
    \includegraphics[width=\linewidth]{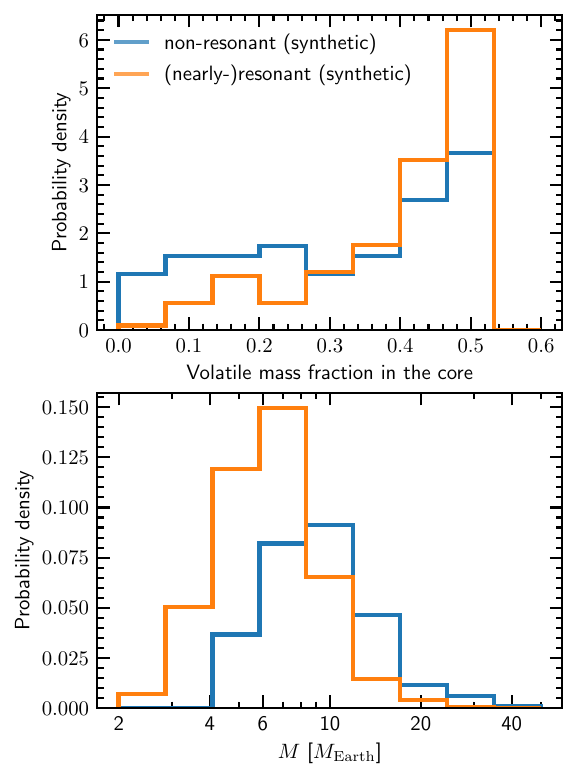}
    \caption{Distributions of synthetic NGPPS planets comparable to the observed sample. Top histogram shows the volatile core mass fraction distribution, while the lower panel shows the total mass of the planets meeting the selection criteria (as in Fig. \ref{fig:NGPPS_a_m_scatter} after 100\,Myr).}
    \label{fig:NGPPS_hists}
\end{figure}

Finally, we briefly discuss the origin of the trends with distance to mean-motion resonance. To make a comparison  with the observed systems, we selected similar-sized, close-in planets, and display their properties in Fig. \ref{fig:models} as well as in Figs. \ref{fig:NGPPS_a_m_scatter} and \ref{fig:NGPPS_hists}. From the latter figures, we can see that more collisions occur after the end of the disk lifetime for the non-resonant population and that their volatile fraction is on average lower. In addition, they also have larger masses and (in order to meet the radius selection criterion) they have higher bulk densities.

\end{appendix}

\end{document}

%% file: fullstrs.txt
\newcommand\pfullm{$2.3e-06_{-2.1e-06}^{+2.7e-05}$}

\newcommand\nbpfull{133}

%% file: controlledstrs.txt
\newcommand\pcontrolledm{$0.002_{-0.001}^{+0.010}$} 
\newcommand\pcontrolledr{$0.49_{-0.24}^{+0.31}$} 
\newcommand\pcontrolledteq{$0.62_{-0.13}^{+0.06}$} 
\newcommand\pcontrolledmet{$0.25_{-0.18}^{+0.31}$} 
\newcommand\pcontrolledteff{$0.62_{-0.18}^{+0.12}$} 
\newcommand\nbpcontrolledres{20} 
\newcommand\nbpcontrollednres{16} 
\newcommand\nbpcontrolled{36} 

%% file: controlled_table.txt
\begin{table*} 
\begin{small} 
\centering 
\caption{\label{tab:controlled} Controlled sample}
\begin{tabular}{llllllll} 
\hline 
planet &  P[day] & M[M$_{Earth}$] & R[R$_{Earth}$]  & $\rho_{rel}$ & $\Delta_{MMR}$ & method & source  \\ 
\hline\hline 
TOI-178 g& $20.717$& $4.40\pm {0.38}$& $2.939\pm {0.056}$& $0.421\pm {0.038}$ &0.001&TTV& \citep{Leleu2024} \\ 
K2-138 e& $8.261$& $12.97\pm {1.99}$& $3.39\pm {0.13}$& $1.00\pm {0.16}$ &0.001&RV& \citep{Lopez2019} \\ 
K2-138 f& $12.758$& $1.63\pm {1.65}$& $2.90\pm {0.14}$& $0.16\pm {0.16}$ &0.001&RV& \citep{Lopez2019} \\ 
Kepler-223 b& $7.385$& $7.40\pm {1.20}$& $2.99\pm {0.23}$& $0.69\pm {0.14}$ &0.001&TTV& \citep{2016Natur.533..509M} \\ 
Kepler-223 c& $9.848$& $5.10\pm {1.40}$& $3.44\pm {0.25}$& $0.39\pm {0.11}$ &0.001&TTV& \citep{2016Natur.533..509M} \\ 
Kepler-26 c& $17.250$& $7.48\pm {0.48}$& $3.11\pm {0.14}$& $0.658\pm {0.062}$ &0.006&TTV& \citep{Leleu2023} \\ 
Kepler-26 b& $12.280$& $4.85\pm {0.43}$& $3.22\pm {0.15}$& $0.406\pm {0.045}$ &0.006&TTV& \citep{Leleu2023} \\ 
Kepler-177 b& $36.855$& $5.84\pm {0.84}$& $3.50\pm {0.17}$& $0.430\pm {0.069}$ &0.007&TTV& \citep{Vissapragada2020} \\ 
K2-266 e& $19.482$& $14.30\pm {5.70}$& $2.73\pm {0.12}$& $1.53\pm {0.62}$ &0.008&TTV & \citep{Rodriguez2018} \\ 
K2-266 d& $14.697$& $8.90\pm {4.75}$& $2.93\pm {0.13}$& $0.86\pm {0.46}$ &0.008&TTV & \citep{Rodriguez2018} \\ 
Kepler-11 d& $22.687$& $6.80\pm {0.75}$& $3.30\pm {0.20}$& $0.547\pm {0.078}$ &0.011&TTV& \citep{HaLi2017} \\ 
Kepler-23 c& $10.740$& $7.81\pm {1.26}$& $3.005\pm {0.074}$& $0.72\pm {0.12}$ &0.012&TTV& \citep{Leleu2023} \\ 
Kepler-305 d& $16.739$& $6.20\pm {1.55}$& $2.76\pm {0.12}$& $0.65\pm {0.17}$ &0.019&TTV& \citep{Leleu2023} \\ 
K2-136 c& $17.307$& $18.10\pm {1.85}$& $3.00\pm {0.13}$& $1.68\pm {0.20}$ &0.022&RV$^*$& \citep{Mayo2023} \\ 
Kepler-57 b& $5.720$& $25.06\pm {5.03}$& $3.135\pm {0.089}$& $2.18\pm {0.45}$ &0.028&TTV& \citep{Leleu2023} \\ 
Kepler-36 c& $16.233$& $7.13\pm {0.18}$& $3.679\pm {0.093}$& $0.488\pm {0.022}$ &0.031&TTV& \citep{Vissapragada2020} \\ 
TOI-125 c& $9.151$& $6.63\pm {0.99}$& $2.759\pm {0.100}$& $0.70\pm {0.11}$ &0.034&RV& \citep{Nielsen2020} \\ 
K2-32 d& $31.717$& $6.70\pm {2.50}$& $3.48\pm {0.12}$& $0.50\pm {0.19}$ &0.035&RV& \citep{Lillo-Box2020} \\ 
K2-32 c& $20.661$& $8.10\pm {2.40}$& $3.13\pm {0.11}$& $0.70\pm {0.21}$ &0.035&RV& \citep{Lillo-Box2020} \\ 
Kepler-33 e& $31.784$& $5.50\pm {1.15}$& $3.50\pm {0.75}$& $0.41\pm {0.15}$ &0.040&TTV& \citep{HaLi2017} \\ 
Kepler-549 b& $42.950$& $11.00\pm {3.70}$& $3.10\pm {0.40}$& $0.97\pm {0.38}$ &0.078&TTV& \citep{HaLi2017} \\ 
Kepler-89 c& $10.424$& $7.80\pm {2.70}$& $3.80\pm {0.65}$& $0.51\pm {0.22}$ &0.143&TTV& \citep{HaLi2017} \\ 
WASP-47 d& $9.031$& $14.20\pm {1.30}$& $3.567\pm {0.045}$& $1.018\pm {0.095}$ &0.171&RV$^*$& \citep{Bryant2022} \\ 
TOI-125 d& $19.980$& $13.60\pm {1.20}$& $2.93\pm {0.17}$& $1.31\pm {0.16}$ &0.183&RV& \citep{Nielsen2020} \\ 
EPIC 249893012 c& $15.624$& $14.67\pm {1.86}$& $3.67\pm {0.15}$& $1.01\pm {0.14}$ &0.288&RV& \citep{Hidalgo2020} \\ 
EPIC 249893012 d& $35.747$& $10.18\pm {2.44}$& $3.94\pm {0.12}$& $0.63\pm {0.15}$ &0.288&RV& \citep{Hidalgo2020} \\ 
HD 136352 c& $27.592$& $11.24\pm {0.64}$& $2.916\pm {0.074}$& $1.088\pm {0.074}$ &0.383&RV& \citep{Delrez2021} \\ 
HD 73583 b& $6.398$& $10.20\pm {3.25}$& $2.790\pm {0.100}$& $1.05\pm {0.34}$ &0.951&RV$^*$& \citep{Barragan2021} \\ 
K2-138 g& $41.968$& $4.32\pm {4.14}$& $3.01\pm {0.28}$& $0.40\pm {0.39}$ &1.290&RV& \citep{Lopez2019} \\ 
EPIC 220674823 c& $13.340$& $8.90\pm {2.40}$& $2.836\pm {0.079}$& $0.90\pm {0.25}$ &21.350&RV& \citep{Bonomo2023} \\ 
TOI-431 d& $12.461$& $9.90\pm {1.51}$& $3.290\pm {0.090}$& $0.80\pm {0.13}$ &23.428&RV& \citep{Osborn2021} \\ 
HD 3167 c& $29.845$& $11.13\pm {0.76}$& $3.00\pm {0.33}$& $1.03\pm {0.18}$ &29.101&RV& \citep{Bonomo2023} \\ 
HD 183579 b& $17.471$& $19.70\pm {3.95}$& $3.55\pm {0.14}$& $1.42\pm {0.30}$ &-&RV& \citep{Palatnick2021} \\ 
TOI-220 b& $10.695$& $13.80\pm {1.00}$& $3.03\pm {0.15}$& $1.26\pm {0.13}$ &-&RV& \citep{Hoyer2021} \\ 
TOI-1052 b& $9.140$& $16.90\pm {1.70}$& $2.87\pm {0.27}$& $1.68\pm {0.29}$ &-&RV& \citep{Armstrong2023} \\ 
TOI-1231 b& $24.246$& $15.40\pm {3.30}$& $3.65\pm {0.15}$& $1.07\pm {0.24}$ &-&RV$^*$& \citep{Burt2021} \\ 
\end{tabular} 
\tablefoot{  
RV planets with no marker have either been only observed by HARPS, or by simultaneously (i.e. overlapping points withing few tens of days) by HARPS and another telescope. K2-136 was first observed by HARPS-N then ESPRESSO, WASP-47 was first observed by Coralie, then by ESPRESSO, HD 73583 was first observed by Coralie, then by HARPS, and TOI-1231 was first observed by PFS.  }
\end{small} 
\end{table*}